\documentclass[conference]{IEEEtran}
\IEEEoverridecommandlockouts

\makeatletter
\let\MYcaption\@makecaption
\makeatother

\usepackage{booktabs} 
\usepackage{amsmath,amssymb,amsfonts}
\usepackage{enumitem}
\usepackage{graphicx}
\usepackage{xcolor}
\usepackage{url}
\usepackage{multirow}
\usepackage{xspace}
\usepackage{subcaption}
\usepackage{balance}
\usepackage{pgf}

\usepackage{hhline}
\usepackage{pifont}
\usepackage[capitalize, noabbrev]{cleveref}

\usepackage[T1]{fontenc}

\makeatletter
\let\@makecaption\MYcaption
\makeatother

\def\BibTeX{{\rm B\kern-.05em{\sc i\kern-.025em b}\kern-.08em
    T\kern-.1667em\lower.7ex\hbox{E}\kern-.125emX}}

\newcommand{\argmax}{\mathop{\mathrm{argmax}}}   
\newcommand{\name}{SIMFL\xspace}
\newcommand{\dfj}{\textsc{Defects4J}\xspace}

\newcommand{\fix}[1]{\textcolor{black}{#1}}

\begin{document}

\title{Ahead of Time Mutation Based Fault Localisation using Statistical Inference}

\author{\IEEEauthorblockN{Jinhan Kim}
\IEEEauthorblockA{\textit{School of Computing} \\
\textit{KAIST} \\
Daejeon, South Korea \\
jinhankim@kaist.ac.kr}
\and
\IEEEauthorblockN{Gabin An}
\IEEEauthorblockA{\textit{School of Computing} \\
\textit{KAIST}\\
Daejeon, South Korea \\
agb94@kaist.ac.kr}
\and
\IEEEauthorblockN{Robert Feldt}
\IEEEauthorblockA{\textit{Dept. of Computer Science and Engineering} \\
\textit{Chalmers University of Technology}\\
Gothenburg, Sweden \\
robert.feldt@chalmers.se}
\and
\IEEEauthorblockN{Shin Yoo}
\IEEEauthorblockA{\textit{School of Computing} \\
\textit{KAIST}\\
Daejeon, South Korea \\
shin.yoo@kaist.ac.kr}
}

\maketitle

\begin{abstract}
    Mutation analysis can effectively capture the dependency between source code 
    and test results. This has been exploited by Mutation Based Fault Localisation 
    (MBFL) techniques. However, MBFL techniques suffer from the need to expend the 
    high cost of mutation analysis after the observation of failures, which may 
    present a challenge for its practical adoption. We introduce \name (Statistical Inference for Mutation-based Fault Localisation), an MBFL 
    technique that allows users to perform the mutation analysis in advance 
    \emph{before} a failure is observed, allowing the amortisation of the analysis 
    cost. \name uses mutants as artificial 
    faults and aims to learn the failure patterns among test cases against 
    different locations of mutations. Once a failure is observed, \name requires 
    either almost no or very small additional cost for analysis, depending on the 
    used inference model. 
    An empirical evaluation using \dfj shows that \name can successfully
    localise up to 113 out of 203 studied faults (55\%)
    at the top, and 159 (78\%) faults within the top five,
    significantly outperforming existing MBFL techniques while using
    the results of mutation analysis
    that has been undertaken before the test failure.
    The amortised cost of mutation analysis can be further
    reduced by mutation sampling: \name retains 80\% of its localisation
    accuracy at the top rank when using only 10\% of generated mutants, 
    compared to results obtained without sampling.
\end{abstract}

\begin{IEEEkeywords}
Fault Localisation
\end{IEEEkeywords}

\section{Introduction}
\label{sec:intro}
As software systems grow in size and complexity, automated fault 
localisation techniques~\cite{Wong:2016aa} have received a lot of attention~\cite{Yoo:2012kx,Sohn:2017xq,Abreu:2011kq,Naish:2011fk,Yoo:2017ss,Wong:2007rt}. 
There are two driving motivations for automated fault localisation. First, 
various studies have shown that developers can benefit from automated fault 
localisation technique if the location of a real fault can be narrowed down to 
a sufficiently small candidate set~\cite{Kochhar:2016aa,Xia2016aa}. Second, 
Automated Program Repair (APR), another technique increasingly in demand,
depends on the accuracy of automated fault localisation for its 
success~\cite{Qi:2013fk,Weimer:2009fk,Wen2018dk}.

Mutation analysis has been successfully applied to fault localisation, 
resulting in a group of techniques called Mutation Based Fault Localisation 
(MBFL)~\cite{Moon:2014ly,Hong:2017qy,Papadakis:2012fk,Papadakis:2015sf,Pearson:2017aa}. Mutation analysis 
applies random syntactic modifications (each corresponding to a mutation operator) to existing code, and observes whether 
the changes in the program behaviour are detected via 
testing~\cite{papadakis2019mt}. Existing MBFL techniques exploit the captured 
dependency between the artificial faults (i.e., mutants) and the changes in 
program behaviours (i.e., test results). For example, if mutating a program 
causes test cases to fail in a pattern similar to an observed failure, the 
mutant may be near the root cause of the observed 
failure~\cite{Papadakis:2012fk,Papadakis:2015sf}. Alternatively, if mutating a 
program causes test cases to fail in a pattern very different from an observed 
failure, the mutant may be far from the location of the root cause~\cite{Moon:2014ly}.

Despite their success, MBFL techniques share a major weakness with mutation 
testing, which is the cost of test execution~\cite{papadakis2019mt}. The more 
closely mutants approximate real faults, the more accurate MBFL techniques 
can be. As such, MBFL benefits from a large number of mutants, generated by a 
diverse set of mutation operators, to be analysed. However, this directly 
increases the cost of inspecting whether each mutant can be \emph{killed} (i.e.,
whether the behavioural differences introduced by them are detectable), 
as this process requires the execution of the test suite per each mutant. 

With large systems, this cost can grow significantly large, to the point that
MBFL techniques cannot be used just-in-time after failures are observed. This 
is especially the case when MBFL techniques are used in the context of 
Continuous Integration (CI)~\cite{duvall2007continuous,meyer2014continuous}. 
If developers encounter a failure during pre-commit testing, they are 
likely to want a just-in-time debugging technique that ensures fast and 
accurate feedback, so that they can remove the fault and continue to submit 
the changes. If, on the other hand, failure is observed during the 
post-commit testing initiated by the CI, it is still crucial 
for a fault localisation technique to be sufficiently fast so that developers 
do not wait hours for feedback~\cite{Liang2018fv}. The cost of having to 
re-run MBFL for each of the possibly many different failure patterns that 
can arise during pre- and post-commit testing efforts over, possibly, several 
commits could be truly staggering.

To overcome the high cost of mutation analysis in MBFL, we introduce \name (Statistical Inference for Mutation-based Fault Localisation), an 
MBFL technique that allows developers to perform the mutation analysis in advance 
against an earlier version of the code. \name constructs a kill matrix 
using a version of the System Under Test (SUT) before any test failures are observed. 
The matrix essentially captures which 
test cases fail when specific locations of SUT are mutated. Once an actual 
failure is observed, \name builds predictive models and consults them using 
the information of which test cases pass and/or fail under the observed 
failure. Depending on the statistical inference technique, 
the actual post-hoc analysis, required after the observation of the 
failure, takes either virtually no time at all (Bayesian inference), or a small 
fraction of mutation analysis time (Logistic
Regression or Multi-Layer Perceptron). \name allows 
developers to amortise the cost of mutation analysis and use MBFL techniques 
in a just-in-time manner. By doing even the model building ahead-of-time the
cost can be amortised further since we only need to use the previously built 
model and apply it to the specific failure patterns that are observed.

We have implemented and evaluated \name using multiple modelling schemes and
statistical inference techniques.
The empirical 
evaluation studies real-world faults in \dfj benchmark~\cite{Just:2014aa}, 
using the Major mutation tool~\cite{Just2011gq}. \name can successfully 
localise up to 113 out of 203 faults at the top, and 159 faults within the 
top five places. To reduce the cost of \name even further, we also
evaluate the impact of mutation sampling on the mutation analysis step of \name. 
When using only 10\% of the generated mutants for analysis, \name can still
achieve 80\% of its localisation accuracy, compared to when not using
sampling. The technical contributions of this paper are as follows:

\begin{itemize}
\item We introduce \name, a Mutation Based Fault Localisation (MBFL)
technique that allows ahead-of-time mutation analysis. Using the outcome of 
the mutation analysis, \name builds a predictive model that allows developers
to predict the location of actual future faults, using the test failure 
information as input. This process significantly amortises the cost of 
mutation analysis.

\item We present the results of an empirical evaluation of \name using the real 
world Java faults in \dfj benchmark. The empirical study concerns not only
the localisation accuracy compared to the state-of-the-art FL techniques,
but also various related aspects of \name such as
the impact of different modelling schemes, the viability of models built
earlier than faults, and the impact of sampling rates.

\item We discuss implications and characteristics of \name and the impact of filtering
mutants by their kill reason. Our observations suggest that mutant filtering has impact
on localisation effectiveness of \name and a potentially effective
hybridisation would be possible between \name and other fault localisation techniques.

\end{itemize}

The rest of the paper is organised as follows. Section~\ref{sec:methodology} lays out 
the foundations of \name by describing how the results of mutation analysis 
are formulated into predictive models for fault localisation. 
Section~\ref{sec:exp_design} presents the details of experimental design, including 
the protocols of the empirical study and research questions.
Section~\ref{sec:result} presents and analyses results, while Section~\ref{sec:discussion}
discusses the results in the wider context of fault 
localisation. Section~\ref{sec:threats} considers potential threats to validity, and 
Section~\ref{sec:related_work} presents related work. Finally, Section~\ref{sec:conclusion} 
concludes and presents future work.

\section{Methodology}
\label{sec:methodology}

Intuitively, the underlying assumption of \name is that,
for a test that has killed the mutants located on a specific program 
element, the same program element should be identified as the suspicious 
location when the same test later fails again. This is based on the coupling effect 
hypothesis in mutation testing: essentially we \emph{simulate} the occurrence
of real faults with artificial faults with known locations, i.e., mutants,
and build predictive models for actual future faults. This section describes
the models and the statistical inference techniques used by \name.

\begin{figure}[ht]
  \centering  
  \includegraphics[width=.4\textwidth]{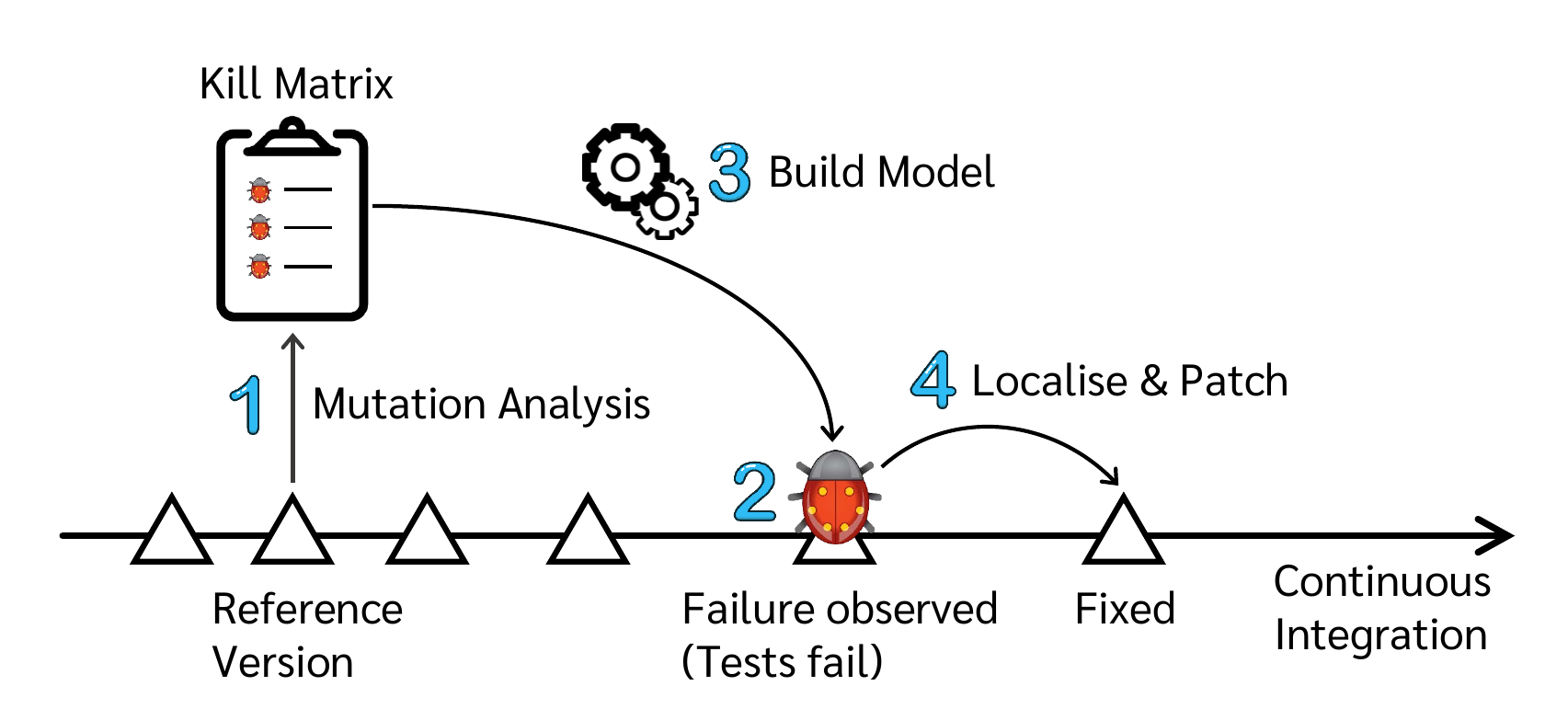}  
  \caption{Expected use case scenario of \name \label{fig:scenario}}  
\end{figure}

Figure~\ref{fig:scenario} depicts the expected use case scenario of \name, 
which includes four stages:

\begin{enumerate}
\item Perform mutation analysis for a version of SUT, and produce the kill matrix. The version is called the \emph{reference} version.
\item While testing a subsequent version, a failure is observed.
\item Using the information of which test case(s) failed, as well as the kill matrix, build a predictive model for fault localisation.
\item Guided by the localisation result, patch the fault.
\end{enumerate}

\subsection{Mutation Analysis}
\label{sec:mutation_analysis}

We perform mutation analysis on the reference version of a program $\mathbf{P}$ with
a test suite $\mathbf{T}$, and compute a kill matrix $\mathbf{K}$,
which contains a complete report of all tests executed on all mutants.
Let $\mathbf{K}_m$ denote a set of tests that kill mutant $m$,
let $\mathbf{X}_e$ be a set of mutants located on a program element $e \in 
\mathbf{P}$,
let $M_e$ be an event that $e$ is mutated, and let $F_t$ be an event that a 
test case $t$ fails on a given program.
Based on the kill matrix $\mathbf{K}_m$, we can approximate the probability of
test case $t$ killing the mutants located on the program element $e$ as follows:

\small
\begin{equation}
  \fix{Pr}\left(F_t \mid M_e \right) \simeq
	\frac{|\left\{m \in \mathbf{X}_e \mid t \in \mathbf{K}_m\right\}|}{|\mathbf{X}_e|}
\label{eq:prob_org}
\end{equation}
\normalsize

  Note that this is strictly an \emph{approximation} based on the observed kill matrix because
  it is impossible to produce and evaluate all possible mutants.
  The exact value of $Pr\left(M_e\right)$ is the ratio of the number of all 
  possible mutants on $e$ to the number of all possible mutants on 
  $\mathbf{P}$; for $\fix{Pr}\left(F_t \mid M_e \right)$, we need to calculate the number of all possible mutants in $e$ that are killed by $t$. Neither
  is feasible. Consequently, we assume that we can analyse
  a \emph{finite} set of mutants that allow us to 
  approximate Equation~\ref{eq:prob_org}.

Next, using Bayes' rule, we calculate the revised probability of the event that 
the program element $e$ has been mutated, given that the test case $t$ fails:

\small
\begin{equation}
  \begin{split}
    \fix{Pr}\left(M_e \mid F_t \right)
    &= \frac{\fix{Pr}\left(F_t \mid M_e \right)\fix{Pr}\left(M_e\right)}{\fix{Pr}\left(F_t\right)} \\
    &\simeq \fix{Pr}\left(\text{fault exists in } e \mid F_t\right)
  \end{split}
\label{eq:prob_bayes}
\end{equation}
\normalsize

We argue that, if real faults are coupled to mutants, the probability above 
can approximate the likelihood that the fault is located on the program 
element $e$, when $t$ is a failing test case in the future. This allows us to 
make ranking models that sort the program elements in descending 
order of the probability.

\subsection{Ranking Models}
\label{sec:ranking_model}

We regard the probability in Equation~\ref{eq:prob_bayes} as the quantitative 
score representing how suspicious the program element $e$ is for the failure observed via
the failure of $t$. This section presents the formulations of ranking models 
based on the scores as well as more refined inference models based on kill matrix data.

\subsubsection{Exact Matching (EM)}
\label{sec:exact_matching}

This model is an extension of Equation~\ref{eq:prob_bayes} to a set of test 
cases. Let $\mathbf{T} = \{t_i \mid 1 \leq i \leq n \leq n'\}$ be the test set,
which consists of two disjoint sets: $\mathbf{T}_f = \{t_1, \dots, t_n\}$ is 
the set of failing test cases, and $\mathbf{T}_p = \mathbf{T} \setminus \mathbf
{T}_f$ is the set of passing tests, on the faulty program. While there can be
many different formulations of ranking models based on a set of test cases, we
start by treating the set of all observed failures, $F_{\mathbf{T}_f}$, as
a conjunctive event of individual test case failures, 
i.e., $F_{\mathbf{T}_f} = F_{t_1} \cap \dots \cap F_{t_n}$. Our goal is to 
find the faulty program element $e_i \in \mathbf{P}$ with the highest probability of being the cause of the observed failure symptoms, that is, 
$\fix{Pr}\left(M_{e_i} \mid F_{\mathbf{T}_f}\right)$. It follows that:

\small
\begin{equation}
  \begin{split}
    \argmax_{i}\fix{Pr}\left(M_e \mid F_{\mathbf{T}_f}\right)
     &= \argmax_{i}\frac{\fix{Pr}\left(F_{\mathbf{T}_f} \mid M_e\right)\fix{Pr}\left(M_e\right)}{\fix{Pr}\left(F_{\mathbf{T}_f}\right)}
  \end{split}  
  \label{eq:prob_bayes_test_set}
\end{equation}
\normalsize

The denominator in Equation~\ref{eq:prob_bayes_test_set}, $\fix{Pr}\left(F_{\mathbf{T}_f}
\right)$, can be ignored without affecting the order of ranking based on this 
score, because it is not related to a specific program element. Expanding the 
numerator yields the following:
\small
\begin{equation}
  \begin{split}
    &\argmax_{i}\fix{Pr}\left(F_{\mathbf{T}_f} \mid M_{e_i}\right)\fix{Pr}\left(M_{e_i}\right) \\
    &= \argmax_{i}\fix{Pr}\left(F_{t_1} \cap \dots \cap F_{t_n} \mid M_{e_i}\right)\fix{Pr}\left(M_{e_i}\right) \\
    &= \argmax_{i}\frac{|\{m \in \mathbf{X}_{e_i} \mid \{t_1, \dots, t_n\} = \mathbf{K}_m\}|}{|\mathbf{X}_{e_i}|}\fix{Pr}\left(M_{e_i}\right) \\
    &= \argmax_{i}\frac{|\{m \in \mathbf{X}_{e_i} \mid \mathbf{T}_f = \mathbf{K}_m\}|}{|\mathbf{X}_{e_i}|}\frac{|\mathbf{X}_{e_i}|}{|\mathbf{X}_{\mathbf{P}}|} \\
    &= \argmax_{i}\frac{|\{m \in \mathbf{X}_{e_i} \mid \mathbf{T}_f = \mathbf{K}_m\}|}{|\mathbf{X}_{\mathbf{P}}|} \\
    &= \argmax_{i}|\{m \in \mathbf{X}_{e_i} \mid \mathbf{T}_f = \mathbf{K}_m\}| 
  \end{split}  
  \label{eq:argmax_only_failing}
\end{equation}
\normalsize

Intuitively, Equation~\ref{eq:argmax_only_failing} counts the mutants on $e$
that cause the same set of test cases to fail as the symptom of the actual 
fault, $F_{\mathbf{T}_f}$. We call this model the Exact Matching (EM) model 
with failing test cases, denoted by EM(F).

Alternatively, we can include passing tests in the pattern matching as well.
Let $P_t$ be an event that a test case $t$ passes on a given program, then
Equation~\ref{eq:argmax_only_failing} changes as follows:

\small
\begin{equation}
  \begin{split}
    &\argmax_{i}\fix{Pr}\left(F_{\mathbf{T}_f} \cap P_{\mathbf{T}_p} \mid M_{e_i}\right)\fix{Pr}\left(M_{e_i}\right) \\
    &= \argmax_{i}\fix{Pr}\left(F_{t_1} \cap \dots \cap F_{t_n} \cap P_{t_{n+1}} \cap \dots \cap P_{t_{n'}} \mid M_{e_i}\right)\fix{Pr}\left(M_{e_i}\right) \\
    &= \argmax_{i}|\{m \in \mathbf{X}_{e_i} \mid \mathbf{T}_f = \mathbf{K}_m \wedge \mathbf{T}_p = \mathbf{T} \setminus \mathbf{K}_m \}|
  \end{split}  
  \label{eq:argmax_with_passing}
\end{equation}
\normalsize

Similarly to EM(F), this model is called EM(F+P): it counts the mutants on $e$
that cause the same set of test cases to fail and pass exactly as the symptom 
of the actual fault. If, for example, a test case $t$ passed under the actual fault, EM(F+P) model will not count any mutants that are killed by $t$.

\subsubsection{Partial Matching (PM)}
\label{sec:partial_matching}
The Exact Matching (EM) models lose any partial matches between the symptom 
and the mutation results. Suppose two test cases, $t_1$ and $t_2$, failed 
under the actual fault, but only $t_1$ killed a mutant on the faulty program 
element, i.e., $ \exists t_1, t_2 \in \mathbf{T}_f, t_1 \in \mathbf{K}_m 
\land t_2 \notin \mathbf{K}_m $. The information that $t_1$ kills a mutant on the location of the fault is lost, simply because $t_2$ failed to do the same. To retrieve this partial information, we propose two additional models based on partial matches: a multiplicative partial match model and an additive partial match model.

\begin{itemize}
    \item PM$^*$(F): Multiplicative Partial Match Model w/ Failing Tests
    \small
    \begin{equation}
        \begin{split}
          &\argmax_{i} \prod_{t \in \mathbf{T}_f} \left(\fix{Pr}\left(M_{e_i} \mid F_{t}\right) + \epsilon\right)\\
          &=\argmax_{i} \prod_{t \in \mathbf{T}_f} \left(|\{m \in \mathbf{X}_{e_i} \mid t \in \mathbf{K}_m\}| + \epsilon\right)
        \end{split}  
        \label{eq:multiplicative}
      \end{equation}
    \normalsize

  \item PM$^+$(F): Additive Partial Match Model w/ Failing Tests
  \small
    \begin{equation}
      \begin{split}
        &\argmax_{i} \sum_{t \in \mathbf{T}_f} \fix{Pr}\left(M_{e_i} \mid F_{t}\right)\\
        &=\argmax_{i} \sum_{t \in \mathbf{T}_f} |\{m \in \mathbf{X}_{e_i} \mid t \in \mathbf{K}_m\}|
      \end{split}  
      \label{eq:additive}
    \end{equation}
    \normalsize
\end{itemize}

Intuitively, instead of counting exact matches, we want to aggregate scores
from the relationship between individual failing test cases and all mutants on 
a specific program element. PM$^*$(F) and PM$^+$(F) respectively aggregate individual
scores by multiplication and addition. Note that the PM$^*$(F) model requires a 
small positive quantity $\epsilon$ to prevent the value of the entire formula
from being zero when there exist one or more terms that evaluate to zero:
the value of $\epsilon$ does not affect the ranking.

Similarly to the case of EM models, we can also include the information of test cases that pass under the actual fault. These two models are called PM$^*$(F+P) and PM$^+$(F+P), and defined as follows:

\begin{itemize}
  \item PM$^*$(F+P): Multiplicative Partial Match Model w/ All Tests 
  \small
  \begin{equation}
    \begin{split}
      &\argmax_{i} \left(\prod_{t \in \mathbf{T}} \left(\fix{Pr}\left(M_{e_i} \mid F_{t}\right) + \epsilon\right) \prod_{t \in \mathbf{T}_p} \left(\fix{Pr}\left(M_{e_i} \mid P_{t}\right) + \epsilon\right)\right)\\
      &=\argmax_{i} \prod_{t \in \mathbf{T}} \left(|\{m \in \mathbf{X}_{e_i} \mid t
      \in \mathbf{T}_f \iff t \in \mathbf{K}_m \}| + \epsilon\right)
    \end{split}
    \label{eq:multiplicative_fp}
  \end{equation}
  \normalsize
  \item PM$^+$(F+P): Additive Partial Match Model w/ All Tests
  \small
  \begin{equation}
    \begin{split}
      &\argmax_{i} \left(\sum_{t \in \mathbf{T}} \left(\fix{Pr}\left(M_{e_i} \mid F_{t}\right)\right) + \sum_{t \in \mathbf{T}_p} \left(\fix{Pr}\left(M_{e_i} \mid P_{t}\right)\right)\right)\\
      &=\argmax_{i} \sum_{t \in \mathbf{T}} |\{m \in \mathbf{X}_{e_i} \mid t \in \mathbf{T}_f \iff t \in \mathbf{K}_m \}|
    \end{split}
    \label{eq:additive_fp}
  \end{equation}
  \normalsize
\end{itemize}

\subsubsection{Linear and Non-linear Classifiers}
\label{sec:classifiers}

Scores from the Bayesian inference models described in Section~\ref{sec:exact_matching}
and~\ref{sec:partial_matching} are directly computed from the kill 
matrix, and requires virtually no additional analysis cost when scores 
are needed to be computed. However, all these models simply rely on counting 
matches between test results under the actual fault and kill matrix from the 
ahead-of-time mutation analysis. 

To investigate if more sophisticated statistical inference techniques can 
improve the accuracy of \name, we apply both linear and non-linear classifiers
to build predictive models. These classifiers take the test results 
as input, and yield the most suspicious method, as well as the suspiciousness score
of each method as output. 
Let $\alpha_\mathbf{T_i}$ denote a 0-1 vector of the test results of 
$\mathbf{T_i}$, where 0 indicates that test case fails, and 1 indicates that 
test case passes. We first build a training set using the kill matrix 
$\mathbf{K}$: test results per mutant $\mathbf{T_i}$ are transformed into $\alpha_\mathbf{T_i}$,
and the class is labelled based on the method where the mutant is located.



We train representative linear and non-linear classifiers using Logistic Regression (LR) 
and Multi-Layer Perceptron (MLP)~\cite{yu2011dual,hinton1990connectionist}.
For our study,
we use a vanilla MLP that consists of one input layer, one hidden layer with 50
neurons, and one output layer. In the serving phase, we use the suspiciousness
score of each program element, which is obtained before the model computes the most
suspicious method. Only using the observed failures,
we can compose 0-1 vectors (i.e., LR(F) and MLP(F)), or compose 0-1 vectors 
by including the information of passing tests (i.e., LR(F+P) and MLP(F+P)). 
Note that, unlike the Bayesian inference models described in  
Section~\ref{sec:exact_matching} and~\ref{sec:partial_matching}, training these 
classifiers requires additional analysis cost to \name, although the training cost of 
these models is much lower than the cost of mutation analysis.


\section{Experimental Design}
\label{sec:exp_design}

This section describes the design of our empirical evaluation, including the 
way we use \dfj benchmark, the research questions, as well as other 
environmental factors.

\subsection{Protocol}
\label{sec:protocol}

One foundational assumption of \name is that existing test cases can be fault revealing
also for future changes. That is, for future faults to which \name will be applied,
test cases that would reveal them are available at the time of
the ahead-of-time mutation analysis. We believe this is a likely scenario mainly
in two contexts: regression faults, which are defined as failures of existing
test cases, and pre-commit testing, for which developers depend on existing test
cases for a sanity check. \name is designed to reduce the cost of MBFL for these
scenarios.\footnote{Although we do note that the more mature a software system is 
and the stronger and more complete its test suite is, the more likely it is that 
these conditions hold and thus that the proposed approach can be useful.}

However, this makes realistic experiments on real-world data challenging since
a majority of failure triggering changes are not likely to have been committed 
to the main branch of the Version Control System (VCS): one of the 
purposes of Continuous Integration is to prevent such commits~\cite{Liang2018fv}. Consequently, 
fault benchmarks, such as \dfj, contain faults that have been reported 
externally (e.g., from issue tracking systems), and provide fault revealing 
test cases that have been added to the VCS with the patch 
itself~\cite{Just:2014aa}. This presents a challenge for the realistic 
evaluation of \name in the context it was designed for. To address 
this issue, we introduce two experimental protocols.

\subsubsection{Faulty Commit Emulation (FCE)}
\label{sec:artificial_scenario}

This scenario emulates a faulty \emph{commit} that would 
trigger failures of existing test cases simply by reversing a fix patch in 
\dfj. We take the fixed version ($V_{fix}$) in \dfj as the reference version and performs the mutation analysis, 
including the test cases from the same version. Subsequently, we reverse the 
fix patch, execute the same test cases, and try to localise the fault using
the results with \name.

We argue that this is more realistic than injecting mutation faults 
artificially to evaluate \name. Since mutants are exactly what \name uses to 
build its models, \name may unfairly benefit if evaluated using mutants as
faults. Instead, we emulate faulty commits using faults that some developers 
actually had introduced in real-world software. Existing work on test data 
generation has also used the fixed version as the reference version, against
which a test generation tool is applied. The reversed fix patch is then
used to emulate regression faults for the evaluation of the generated 
tests~\cite{Shamshiri2015ase,Just2014fse}. Our approach with FCE
is similar in the sense that we analyse the fixed version first,
then use the outcome to localise the emulated regression fault.


\subsubsection{Test Existence Emulation (TEE)}
\label{sec:realworld_scenario}

This scenario uses original faulty commits that led to the faulty versions 
($V_{bug}$) in \dfj, but simply \emph{pretends} that the \emph{fault revealing test cases
existed earlier}. We have checked whether the fault revealing test cases in \dfj
can be executed against versions that precede the actual faulty version.
Since system specifications evolve over time, executing a future test case
against past versions is not always successful: we have identified 28 previous
versions for which the future fault revealing test cases can be executed and 
\emph{do not fail}. We use these 28 versions as references, and use their
mutation analysis results to localise the corresponding faults that happened later.
Compared to FCE, 
TEE follows the ground truth code changes, and only assumes the earlier 
existence of fault revealing test cases. We use TEE to complement the FCE 
scenario. Specifically, TEE can evaluate whether training \name models with 
kill matrices of earlier versions degrades its 
localisation accuracy.

\subsubsection{Experimental Premise}
\label{sec:experimental_premise}

Building a full kill matrix requires huge computational cost: mutation 
analysis on all versions of Closure using Major exceeded our 24 hours 
timeout, and other subject programs also required significant amounts of
analysis time. To address this practical concern, for empirical evaluation,
we have constructed the kill matrix using only the 
\emph{relevant test cases} as defined by \dfj\footnote{See https://github.com/rjust/defects4j/tree/v1.3.1\#export-version-specific-properties}, 
which include the failing test cases as well as any passing test cases that 
makes the JVM to load at least one of the classes modified by the fault 
introducing commit.

Note that this procedure has been adopted strictly to reduce experimental cost.
Since we only have the kill matrix for the relevant test cases, models that 
use F+P test cases actually use the full set of relevant test cases. However, 
if construction of the full kill matrix is feasible, the same input used by 
\name in this paper is naturally available. The F+P models can be trained  
either using the full set of test cases (increased training cost but also
richer input information), or using the relevant test cases (relevancy
information is still cheaper than full coverage instrumentation).
We argue that, in general, the limitation to only the relevant test cases is a
conservative one and should reduce rather than improve the fault localisation
accuracy of \name since other test cases could also be informative for its
statistical models.

\subsubsection{Using Test Runtime Information}
\label{sec:runtime_info}

The use case of \name assumes that, while the actual mutation analysis can be
performed in advance, the inference models are trained after the observation 
of a failure (see \cref{fig:scenario}). In practice, the observation of the 
behaviour of the failing test cases can provide information that is beyond
the mutation analysis. Consequently, we exploit this additional information 
by collecting coverage reports of failing test cases using \texttt{Cobertura}.
We then exclude any methods and mutants that are not covered by the failing
test cases from model training and the final ranking.

\begin{table}[ht]
  \centering
  \caption{Subject Programs in \dfj \label{tab:subject}}
  \scalebox{0.8}{
  \begin{tabular}{l|rrrrr}
  \toprule
  Subject & \# Faults & kLoC & \# Methods & \# Mutants & \# Test cases\\
  \midrule
  Commons-lang (Lang) & 65  & 50  & 1,527  & 21,178  & 2,245  \\
  JFreeChart (Chart)         & 26  & 132  & 4,903  & 75,985  & 2,205  \\
  Joda-Time (Time)           & 27  & 105  & 1,946  & 21,689  & 4,130  \\
  Closure compiler (Closure) & 133 & 216  & 5,038  & 58,515  & 7,927  \\
  Commons-math (Math) & 106 & 104  & 2,713  & 79,428  & 3,602  \\
  \midrule
  Total                      & 357 & 607 & 16,126 & 256,792 & 20,109 \\
  \bottomrule
  \end{tabular}
  }
\end{table} 

\subsection{Subject Programs}
\label{sec:subject}

In our study, we use 357 versions of five different programs from the \dfj version 1.3.1.
They provide reproducible and isolated faults of real-world
programs.
Table~\ref{tab:subject} summarises the subject programs we used with
the average number of generated mutants, methods, lines of code, and test cases
across all faults belonging to each subject respectively.
We could not include
Mockito as we failed to compile the majority of its versions 
and their mutants using the build script provided by \dfj on Docker containers.




\subsection{Research Questions}
\label{sec:rqs}

\vspace{0.5em}

\noindent\textbf{RQ1. Localisation Effectiveness:} \textit{Does the models of \name produce
	      accurate
        fault localisation compared to the state-of-the-art FL techniques?}
        RQ1 is answered by computing the standard
	      evaluation metrics on the eight models of \name under the FCE scenario
        outlined in~Section~\ref{sec:artificial_scenario}. We compare 
        \name with
        two MBFL techniques (MUSE and Metallaxis), two SBFL
        techniques (Ochiai and DStar), and two learning-to-rank based
        FL techniques (TraPT and FLUCCS).
        
\vspace{0.5em}

\noindent\textbf{RQ2. Model Viability:} \textit{How well does \name hold
	      up when applied using models built earlier?}
        RQ2 is answered by computing
	      the standard evaluation metrics using prior models built under the TEE
	      scenario outlined
        in~Section~\ref{sec:realworld_scenario}.
        
\vspace{0.5em}

\noindent\textbf{RQ3. Sampling Impact:} \textit{What is the impact of
	      mutation sampling to the effectiveness of \name?} Since the cost of mutation
	      analysis is the major component of the cost of \name, we investigate how much
        impact different mutation sampling rates have. We evaluate two different sampling techniques: uniform random sampling, which samples from the pool of all
        mutants uniformly, and stratified sampling, which samples as the equal number of mutants from each method as possible.

\subsection{Evaluation Metrics and Tie Breaking}
\label{sec:metrics}
We use three standard evaluation metrics:

\begin{itemize}
  \item $acc@n$: counts the number of faults located within top $n$ ranks.
                We report $acc@1$, $acc@3$, $acc@5$, and $acc@10$. If a fault is patched across multiple methods, we take the highest ranked method to compute $acc@n$.

  \item \noindent$wef$: approximates the amount of efforts wasted by developer
  while investigating non-faulty methods that are ranked higher than the faulty method.

  \item \noindent Mean Average Precision (MAP): measures the mean of the average
  precision values for a group of all faults.
\end{itemize}

If multiple program elements have the same score, resulting in the same rank, we break the tie using max tie breaker that places all program elements with the same score at the lowest rank.


\begin{table*}[ht]
  \centering
  \caption{Effectiveness of \name models using FCE scenario.\label{tab:RQ1_total_ranks}}
  \scalebox{0.8}{
    \begin{tabular}{l|lr|rrrr|r|r||l|lr|rrrr|r|r}
      \toprule
      Model & Project  & Total & \multicolumn{4}{c|}{$acc$} & $wef$ & MAP & Model & Project  & Total & \multicolumn{4}{c|}{$acc$} & $wef$ & MAP\\
      &   & Studied & @1 & @3 & @5 & @10 & med  & & &  & Studied & @1 & @3 & @5 & @10 & med & \\
      \midrule
      \multirow{6}{*}{\shortstack[l]{EM\\(F)}} & Lang & 62 (65) & 35 & 45 & 47 & 48 & \textbf{0.0} &  0.6176 & \multirow{6}{*}{\shortstack[l]{EM\\(F+P)}} & Lang & 61 (65) & 36 & 41 & 43 & 44 & \textbf{0.0} & 0.5922 \\
      & Chart & 26 (26) & \textbf{6} & \textbf{11} & \textbf{13} & 15 & 5.0 &   0.3294 & & Chart & 25 (26) & 6 & 9 & 10 & 11 & 27.0 & 0.2917 \\
      & Time & 26 (27) & \textbf{4} & 9 & 9 & 13 & 8.5 &  0.2451 & & Time & 26 (27) & 10 & 13 & 14 & 15 & 3.0 & 0.3819 \\
      & Closure & 132 (133) & 10 & 31 & 41 & 57 & 17.0 &   0.1753 & & Closure & 0 (133) & - & - & - & - & - & - \\
      & Math & 102 (106) & 22 & 43 & 53 & 71 & 4.0 & 0.3404 & & Math & 91 (106) & 32 & 45 & 47 & 49 & 3.0  & 0.4098 \\ \cmidrule{2-9} \cmidrule{11-18} 
      & Total & 348 (357) & 77 & 139 & 163 & 204 & & & & Total & 203 (357) & 84 & 108 & 114 & 119 & & \\
      \midrule
      \multirow{6}{*}{\shortstack[l]{PM$^*$\\(F)}} & Lang & 62 (65) & 38 & 47 & 51 & 53 & \textbf{0.0} & 0.6732 & \multirow{6}{*}{\shortstack[l]{PM$^*$\\(F+P)}} & Lang & 61 (65) & 27 & 36 & 37 & 42 & 1.0  & 0.5264 \\
      & Chart & 26 (26) & \textbf{6} & \textbf{11} & \textbf{13} & 16 & 5.0 & 0.3562 & & Chart & 25 (26) & 7 & 9 & 12 & 14 & 6.0  & 0.3598 \\
      & Time & 26 (27) & \textbf{4} & \textbf{10} & 10 & 13 & 8.0  & 0.2549 & & Time & 26 (27) & 1 & 3 & 4 & 12 & 16.0  & 0.1172 \\
      & Closure & 132 (133) & 11 & 36 & 50 & 66 & 9.5  & 0.1982 & & Closure & 0 (133) & - & - & - & - & - & -\\
      & Math & 102 (106) & 23 & \textbf{47} & 59 & 77 & 3.5 & 0.3753 & & Math & 91 (106) & 14 & 26 & 33 & 42 & 12.0 & 0.2460 \\ \cmidrule{2-9} \cmidrule{11-18} 
      & Total & 348 (357) & 82 & 151 & 183 & 225 & & & & Total & 203 (357) & 49 & 74 & 86 & 110 & & \\
      \midrule
      \multirow{6}{*}{\shortstack[l]{PM$^+$\\(F)}} & Lang & 62 (65) & 40 & 48 & 52 & 53 & \textbf{0.0} & 0.6977 & \multirow{6}{*}{\shortstack[l]{PM$^+$\\(F+P)}} & Lang & 61 (65) & 19 & 31 & 33 & 37 & 2.0 & 0.4291 \\
      & Chart & 26 (26) & \textbf{6} & 10 & \textbf{13} & \textbf{19} & \textbf{4.0}  & \textbf{0.3697} & & Chart & 25 (26) & 5 & 9 & 12 & 13 & 8.0  & 0.2712 \\
      & Time & 26 (27) & \textbf{4} & \textbf{10} & 10 & 13 & 8.0  & 0.2564 & & Time & 26 (27) & 0 & 2 & 3 & 5 & 40.5  & 0.0616 \\
      & Closure & 132 (133) & \textbf{12} & \textbf{41} & \textbf{52} & 65 & 11.0  & 0.2005 & & Closure & 0 (133) & - & - & - & - & - & -\\
      & Math & 102 (106) & 24 & 46 & 59 & 77 & 4.0  & 0.3845 & & Math & 91 (106) & 9 & 15 & 20 & 29 & 23.0  & 0.1574 \\ \cmidrule{2-9} \cmidrule{11-18} 
      & Total & 348 (357) & 86 & \textbf{155} & \textbf{186} & \textbf{227} & & & & Total & 203 (357) & 33 & 57 & 68 & 84 & & \\  
      \midrule
      \multirow{6}{*}{\shortstack[l]{LR\\(F)}} & Lang & 62 (65) & \textbf{41} & 49 & \textbf{53} & \textbf{55} & \textbf{0.0}  & \textbf{0.7179} & \multirow{6}{*}{\shortstack[l]{LR\\(F+P)}} & Lang & 61 (65) & 40 & 49 & 51 & 53 & \textbf{0.0} & 0.7017 \\
      & Chart & 26 (26) & 5 & 9 & 12 & 14 & 6.0 & 0.3175 & & Chart & 25 (26) & 8 & 14 & 14 & 16 & \textbf{2.0}  & 0.4194 \\
      & Time & 26 (27) & \textbf{4} & \textbf{10} & \textbf{12} & \textbf{14} & 5.5  & 0.2668 & & Time & 26 (27) & 8 & 14 & 17 & 19 & 2.0 & 0.4094 \\
      & Closure & 132 (133) & \textbf{12} & 37 & 50 & \textbf{68} & \textbf{9.0}  & \textbf{0.2074} & & Closure & 0 (133) & - & - & - & - & - & -\\
      & Math & 102 (106) & \textbf{28} & \textbf{47} & 59 & 75 & \textbf{3.0} & \textbf{0.3976} & & Math & 91 (106) & 32 & 43 & 47 & 51 & 3.0 & 0.4066 \\ \cmidrule{2-9} \cmidrule{11-18} 
      & Total & 348 (357) & \textbf{90} & 152 & \textbf{186} & 226 & & & & Total & 203 (357) & 88 & 120 & 129 & 139 & &  \\
      \midrule
      \multirow{6}{*}{\shortstack[l]{MLP\\(F)}} & Lang & 62 (65) & 39 & \textbf{51} & \textbf{53} & \textbf{55} & \textbf{0.0} & 0.7052 & \multirow{6}{*}{\shortstack[l]{MLP\\(F+P)}} & Lang & 61 (65) & \textbf{48} & \textbf{55} & \textbf{56} & \textbf{56} & \textbf{0.0}  & \textbf{0.7882} \\
      & Chart & 26 (26) & 5 & 10 & 12 & 15 & 6.0 & 0.3319 & & Chart & 25 (26) & \textbf{9} & \textbf{13} & \textbf{15} & \textbf{19} & \textbf{2.0}  & \textbf{0.4477} \\
      & Time & 26 (27) & \textbf{4} & \textbf{10} & \textbf{12} & \textbf{14} & \textbf{5.0} & \textbf{0.2710} & & Time & 26 (27) & \textbf{11} & \textbf{16} & \textbf{18} & \textbf{24} & \textbf{1.0} & \textbf{0.4847} \\
      & Closure & 132 (133) & 11 & 33 & 41 & 60 & 12.0  & 0.1888 & & Closure & 0 (133) & - & - & - & - & - & - \\
      & Math & 102 (106) & 26 & 46 & \textbf{62} & \textbf{79} & \textbf{3.0} & 0.3941 & & Math & 91 (106) & \textbf{45} & \textbf{61} & \textbf{70} & \textbf{82} & \textbf{1.0}  & \textbf{0.5194} \\ \cmidrule{2-9} \cmidrule{11-18} 
      & Total & 348 (357) & 85 & 150 & 180 & 223 & & & & Total & 203 (357) & \textbf{113} & \textbf{145} & \textbf{159} & \textbf{181} & & \\
      \bottomrule
      \end{tabular}
  }
\end{table*} 

\subsection{Mutation Tool and Operators}
\label{sec:mutation_tool_and_operator}

In the study, we use Major version 1.3.4~\cite{Just2011gq} as our mutation analysis tool,
and choose all mutation operators in Major.
Note that some operators had to be turned off for specific classes so
that Major does not generate an exceptionally large number of mutants.\footnote{Due to the internal design of Major, some classes that
yield too many mutants may lead to the violation of bytecode length limit 
imposed by Java compiler. See \url{https://github.com/rjust/defects4j/issues/62} 
for technical details.}



\section{Results}
\label{sec:result}

Due to a space limit, we present the full results including all evaluation metrics
online at \url{https://coinse.github.io/simfl-results}.

\subsection{Effectiveness (RQ1)}
\label{sec:result_RQ1}

We start by comparing different \name models. Subsequently, using the best 
\name model, we compare \name to the state-of-the-art fault localisation 
techniques.

\subsubsection{Comparison Between \name Models}

Table~\ref{tab:RQ1_total_ranks} shows the results of each evaluation metric 
for all studied faults, following the FCE scenario. The numbers $X(Y)$ in the 
column "Total Studied" represent the number of faults that we can localise 
$(X)$, and the number of faults provided by \dfj $(Y)$. Evaluation metric values 
representing the
best outcome (i.e., the largest $acc@n$ and MAP, and the smallest $wef$) are typeset
in bold. See 
Section~\ref{sec:subject} for the details of exclusion criteria we used: note 
that more faults are excluded from the study of F+P models shown on the right.

Overall, MLP(F+P) shows the best performance in terms of $acc@n$ metrics,
placing 48 out of 61 faults at the first place for Lang, and 45 out of 91 
faults at the first place for Math. Considering that MLP(F+P) is evaluated
on fewer faults (203) than MLP(F) (348), the result suggests that MLP(F+P)
shows better performance on average.

We argue that including results of 
passing tests gives richer information when compared to only using results of 
failing tests. However, we also note that only MLP significantly benefits from
the additional information: MLP(F+P) places 28 more faults at the top than 
MLP(F). Two linear models, LR(F) and LR(F+P), on the other hand, do not show 
any significant difference in performance. This suggests that exploiting this 
information requires more sophisticated, non-linear inference methods. 

The reason that PM$^+$(F) shows comparable results to MLP(F) may be that
it is relatively easy to simply count the matching patterns of failing tests, 
which are much rarer than passing tests. We also note that PM$^*$(F) and 
PM$^+$(F) both produce better results than EM(F), suggesting that partial
matches are better than exact matches. This is because even the fault revealing
test case may not be able to kill all mutations applied to the location of
the fault. In such a case, 
the EM(F) model will lose the information, while the PM(F) models will benefit from
other killed mutants from the same location.

Finally, the addition of passing test information to PM models actually
degrades the performance significantly, as the metrics for PM$^*$(F+P) and 
PM$^+$(F+P) show. Partially matching test cases that did not fail against the 
faulty version with test cases that did not kill mutants at the location of 
the fault will directly dilute the signal, as failing tests and killed mutants 
are likely to provide more information about the location of the fault in 
general.


\begin{figure}[ht]
    \centering    
    \includegraphics[width=.45\textwidth]{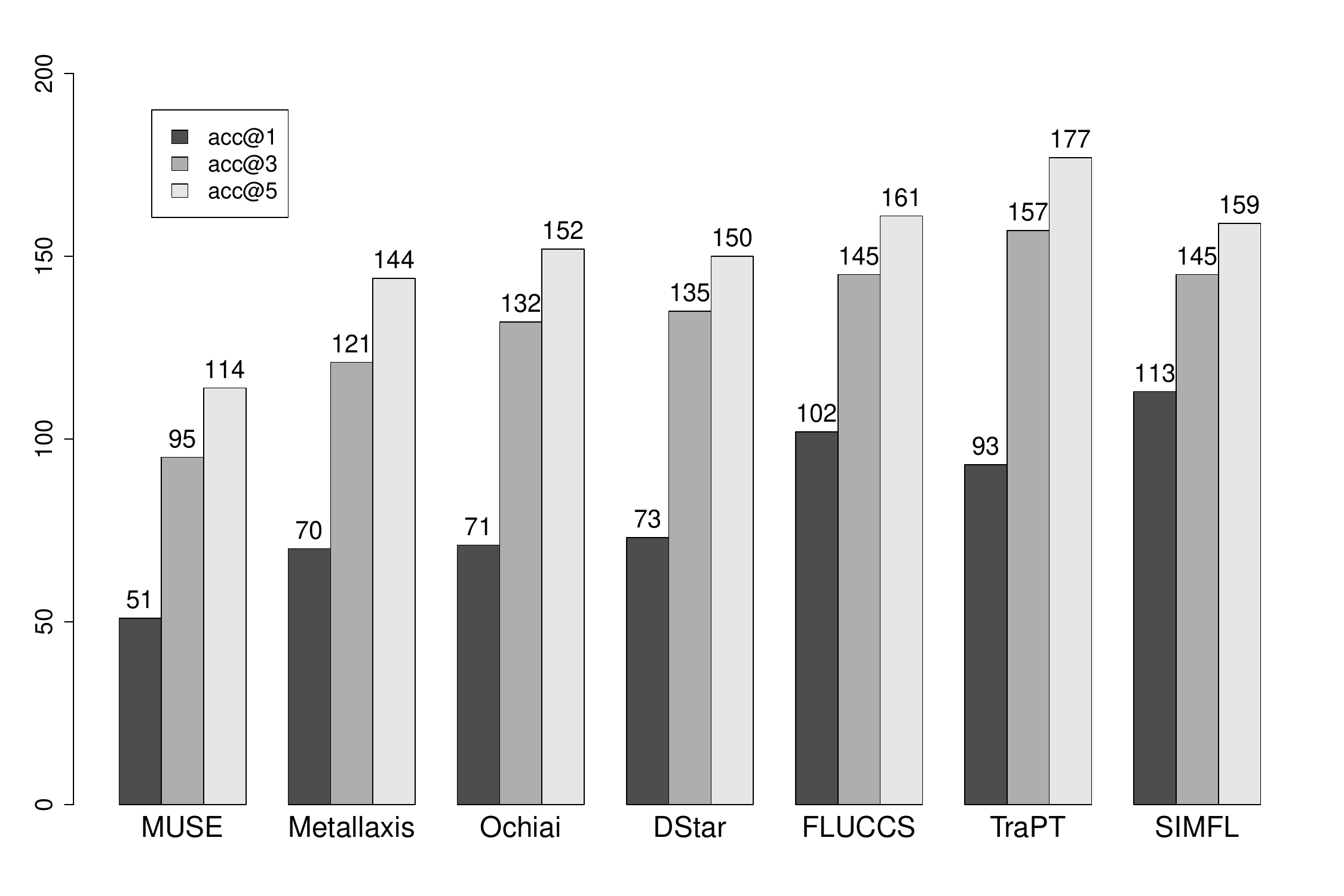}
    \caption{Comparison to other FL techniques:
    $acc@n$ metric values without counts of Closure\label{fig:RQX_rank_comparison}}
\end{figure}

\subsubsection{Comparison to Other FL Techniques}
\label{sec:comparison}

To gain some insights into the trade-off between amortised modelling efforts 
and localisation accuracy, we compare the method-level
fault localisation results of the state-of-the-art MBFL and SBFL
techniques, the result of which is shown in \cref{fig:RQX_rank_comparison}. 
We obtained the performance of the each model (i.e., $acc@n$) on \dfj
from the literatures, and artefact of Zou et al.~\cite{zou2019empirical}.
Based on the results of the comparison between \name models, we choose
MLP(F+P) to represent \name. However, since Closure has been excluded from 
the evaluation of F+P models, we have excluded Closure from the results of other techniques for a fair comparison.

\cref{fig:RQX_rank_comparison} shows that MLP(F+P) is better than other 
techniques in terms of $acc@1$, but TraPT performs better in terms of $acc@3$ and $acc@5$.
Although \name does not make use of learning-to-rank technique to boost performance
by fully including runtime information or suspiciousness scores of other FL techniques,
\name localises faults at the top better than others, and shows
comparable results to the learning-to-rank techniques: FLUCCS and TraPT.

Based on this analysis, we answer RQ1 that \name can localise faults 
accurately compared to the existing techniques: \name places up to 25.86\%
(90 of 348 for LR(F)) of studied faults at the top using F models,
and 55.67\% (113 of 203 for MLP(F+P)) of studied faults at the top using F+P models.


\subsection{Model Viability (RQ2)}
\label{sec:result_RQ2}

\begin{table}[ht]
  \centering
  \caption{Viability of F models using TEE scenario. The ranks that do not have same ranks with FCE are typeset in bold. \label{tab:RQ2}}
  \scalebox{0.55}{
  \begin{tabular}{ll|lp{0.5cm}p{0.5cm}p{0.5cm}p{0.5cm}p{0.5cm}|ll|lp{0.5cm}p{0.5cm}p{0.5cm}p{0.5cm}p{0.5cm}}
  \toprule
  \multicolumn{2}{l|}{Fault}    & Commit             & \multicolumn{5}{c|}{Rank}                                                            & \multicolumn{2}{l|}{Fault} & Commit             & \multicolumn{5}{c}{Rank} \\
  & & ($\Delta$rev.) & EM                    & PM$^*$   & PM$^+$   & LR                    &  MLP        & & & ($\Delta$rev.) & EM       & PM$^*$   & PM$^+$   & LR       & MLP      \\ \midrule
  \parbox[t]{2mm}{\multirow{18}{*}{\rotatebox[origin=c]{90}{Closure}}} & 21  & FCE Rank           & 2                        & 2           & 2           & 2                        &  2           & \parbox[t]{2mm}{\multirow{18}{*}{\rotatebox[origin=c]{90}{Math}}} & 46  & FCE Rank           & 188         & 188         & 188         & 47          & 85          \\
          &     & 32a12ba (2)        & 2                        & 2           & 2           & 2                        &  2           &      &     & bbb5e1e (1)        & 188         & 188         & 188         & \textbf{35} & \textbf{47} \\
          &     & 43a5523 (3)        & 2                        & 2           & 2           & 2                        &  2           &      &     & 37680e2 (2)        & 188         & 188         & 188         & \textbf{35} & \textbf{47} \\ \cmidrule{2-8}
          & 61  & FCE Rank           & 7                        & 4           & 5           & 5                        &  6           &      &     & 1861674 (3)        & 188         & 188         & 188         & \textbf{35} & \textbf{27} \\ 
          &     & f5529dd (3)        & 7                        & 4           & 5           & 5                        &  6           &      &     & f0b12de (4)        & 188         & \textbf{1}  & \textbf{1}  & \textbf{3}  & \textbf{1}  \\
          &     & b12d1d6 (4)        & 7                        & 4           & 5           & 5                        &  6           &      &     & 8581b76 (5)        & 188         & 188         & 188         & \textbf{35} & \textbf{41} \\ \cmidrule{10-16}
          &     & 245362a (7)        & 7                        & 4           & 5           & 5                        &  \textbf{7}  &  & 89  & FCE Rank           & 13          & 13          & 13          & 8           & 7           \\
          &     & 8abd1d9 (8)        & 7                        & 4           & 5           & 5                        &  \textbf{8}  &      &     & 43336b0 (1)        & \textbf{12} & \textbf{12} & \textbf{12} & \textbf{2}  & \textbf{3}  \\
          &     & 37b0e1b (9)        & 7                        & 4           & 5           & 5                        &  6           &      &     & cdd62a0 (2)        & \textbf{14} & \textbf{14} & \textbf{14} & \textbf{2}  & \textbf{5}  \\ \cmidrule{2-8}
          & 62  & FCE Rank           & 1                        & 1           & 1           & 1                        &  1           &      &     & 90439e5 (3)        & 13          & 13          & 13          & 8           & \textbf{11} \\
          &     & 245362a (2)        & 1                        & 1           & 1           & 1                        &  1           &      &     & 36a8485 (4)        & 13          & 13          & 13          & 8           & \textbf{13} \\
          &     & 8abd1d9 (3)        & 1                        & 1           & 1           & 1                        &  1           &      &     & dbe7842 (5)        & 13          & 13          & 13          & 8           & 7           \\
          &     & 37b0e1b (4)        & 1                        & 1           & 1           & 1                        &  1           &      &     & d84a587 (6)        & 13          & 13          & 13          & 8           & \textbf{12} \\\cmidrule{2-8}
          & 115 & FCE Rank           & 14                       & 22          & 24          & 19                       &  11          &      &     & d27e072 (7)        & 13          & 13          & 13          & 8           & \textbf{10} \\
          &     & b9262dc (5)        & 14                       & \textbf{19} & \textbf{22} & \textbf{18}              &  \textbf{13} &      &     & 3590bdc (8)        & 13          & 13          & 13          & 8           & \textbf{8}  \\
          &     & 911b2d6 (6)        & 14                       & 22          & 24          & 19                       &  \textbf{12} &      &     & 6b108c0 (9)        & 13          & 13          & 13          & 8           & \textbf{13} \\\cmidrule{2-8}
          & 120 & FCE Rank           & 7                        & 7           & 7           & 6                        &  6           &      &     & 9c55428 (10)       & 13          & 13          & 13          & 8           & \textbf{12} \\
          &     & 2aee36e (3)        & \textbf{24}              & \textbf{24} & \textbf{24} & \textbf{15}              &  \textbf{16} &      &     &                    &             &             &             &             &             \\

  \bottomrule
  \end{tabular}}
\end{table}

Following the TEE scenario described in Section~\ref{sec:realworld_scenario},
we seek reference versions preceding the faulty version, i.e., the versions
before the faulty version that pass all test cases
of the fixed program, including the fault revealing test cases. Assuming that
more recent versions are more likely to serve as references, given a faulty version
$n$, we check $n-1$, \dots, $n-10$, $n-20$, and $n-30$ previous program versions,
as it is impractical to inspect all of them. Starting from 357 faulty versions of
subject programs, we found 28 preceding reference versions 
that correspond to seven different faulty versions. We have trained 
five F models on each of the 28 reference versions to localise the fault in the faulty version, resulting in 140 rankings based on TEE scenario.
Note that we did not consider F+P models on these reference versions because they require
more than 24 hours for mutation analysis, as described in Section~\ref{sec:subject}.

Table~\ref{tab:RQ2} shows the rank of the faulty method for each F model
built on each preceding reference version. Out of 140 TEE based rankings produced by F models, 103 are identical to the corresponding FCE ranking. One notable exception
is Math 46 (f0b12de) that shows a significant improvement over the FCE scenario rank. 
We have manually examined the kill matrix of this reference version, and found
that some mutants in the future faulty method have been additionally killed 
due to timeout (enforced by Major itself), contributing to the high rank
(these mutants were not killed
in other preceding reference versions of Math 46). We suspect that this is due 
to the non-determinism in the process of building the kill matrix: the mutation
may have brought in flakiness that has been removed for the original program.
We study the impact of different kill reasons in Section~\ref{sec:kill_reason_filtering},
and furthermore discuss this as one of the threats to internal validity in \cref{sec:threats}.

We answer RQ2 that performances of \name using models built with preceding 
reference versions tend to be stable when compared to the FCE results: only 19
out of 140 cases show degraded performance since we
used less recent mutation analysis results.

\begin{table}[ht]
  \centering
  \caption{Uniform and stratified random sampling\label{tab:RQ3}}
  \scalebox{0.8}{
  \begin{tabular}{l|lr|rr||l|rr}
  \toprule
  Ratio & Model & Total & \multicolumn{2}{c||}{$acc$} & N & \multicolumn{2}{c}{$acc$} \\
  &   & Studied & @1 & @3 & (Ratio) & @1 & @3  \\
  \midrule
  \multirow{10}{*}{\shortstack[l]{0.1}} & EM(F) & 348 & 59.80 & 95.35 &  \multirow{10}{*}{\shortstack[l]{5 \\ (0.27)}} & 36.50 & 65.05  \\
  & PM$^*$(F) & 348 & 66.55 & 107.85 &  & 40.95 & 74.75  \\
  & PM$^+$(F) & 348 & 68.40 & 108.50 &   & 40.05 & 75.00  \\
  & LR(F) & 348 & 71.75 & 121.50 &   & 49.15 & 91.10   \\
  & MLP(F) & 348 & 76.70 & 120.05 &   & 49.90 & 88.85   \\
  & EM(F+P) & 203 & 46.35 & 56.55 & & 39.50 & 57.05  \\
  & PM$^*$(F+P) & 203 & 45.45 & 66.65 & & 65.70 & 100.25   \\
  & PM$^+$(F+P) & 203 & 29.35 & 52.45 & & 39.55 & 52.45  \\
  & LR(F+P) & 203 & 70.70 & 93.30 &   & 75.95 & 114.40   \\
  & MLP(F+P) & 203 & 83.60 & 111.30 &   & 78.15 & 118.20   \\
  \midrule
  \multirow{10}{*}{\shortstack[l]{0.3}} & EM(F) & 348 & 72.25 & 118.55 &\multirow{10}{*}{\shortstack[l]{10 \\ (0.41)}} & 45.60 & 80.95   \\
  & PM$^*$(F) & 348 & 78.90 & 132.15 &   & 49.65 & 93.50   \\
  & PM$^+$(F) & 348 & 83.45 & 133.70 &    & 53.65 & 93.00   \\
  & LR(F) & 348 & 84.85 & 142.15 &    & 59.95 & 106.60   \\
  & MLP(F) & 348 & 82.45 & 139.40 &    & 56.90 & 102.60   \\
  & EM(F+P) & 203 & 66.70 & 82.35 &   & 50.60 & 74.00  \\
  & PM$^*$(F+P) & 203 & 47.80 & 71.40 &   & 74.45 & 106.35   \\
  & PM$^+$(F+P) & 203 & 31.75 & 55.45 &  & 39.20 & 55.05  \\
  & LR(F+P) & 203 & 81.95 & 107.60 &   & 82.15 & 115.90   \\
  & MLP(F+P) & 203 & 101.55 & 132.20 &   & 89.70 & 126.35   \\
  \midrule
  \multirow{10}{*}{\shortstack[l]{0.5}} & EM(F) & 348 & 75.90 & 128.30 & \multirow{10}{*}{\shortstack[l]{15 \\ (0.50)}} & 53.60 & 96.70  \\
  & PM$^*$(F) & 348 & 82.90 & 141.80 & & 58.70 & 110.10  \\
  & PM$^+$(F) & 348 & 86.30 & 143.80 & & 62.20 & 109.15  \\
  & LR(F) & 348 & 88.75 & 145.90 & & 66.70 & 118.30  \\
  & MLP(F) & 348 & 84.70 & 146.00 & & 63.05 & 114.25  \\
  & EM(F+P) & 203 & 73.90 & 93.10 & & 57.35 & 82.65  \\
  & PM$^*$(F+P) & 203 & 47.55 & 72.70 &  & 78.75 & 109.90  \\
  & PM$^+$(F+P) & 203 & 32.70 & 57.00 & & \textbf{42.15} & 61.45 \\
  & LR(F+P) & 203 & 86.35 & 112.15 & & 86.95 & 116.95  \\
  & MLP(F+P) & 203 & 104.90 & 138.65 & & 94.80 & 134.60  \\
  \midrule
  \multirow{10}{*}{\shortstack[l]{0.7}} & EM(F) & 348 & \textbf{78.05} & 133.55  & \multirow{10}{*}{\shortstack[l]{20 \\ (0.56)}} & 55.80 & 105.80 \\
  & PM$^*$(F) & 348 & \textbf{84.80} & 147.75 &  & 64.95 & 121.45 \\
  & PM$^+$(F) & 348 & \textbf{88.15} & 150.05 &  & 70.05 & 124.10 \\
  & LR(F) & 348 & 89.65 & 148.50 &  & 74.35 & 127.05 \\
  & MLP(F) & 348 & 84.30 & 144.60 &  & 69.70 & 124.70 \\
  & EM(F+P) & 203 & 78.05 & 98.85 &  & 62.50 & 88.05 \\
  & PM$^*$(F+P) & 203 & 48.30 & 73.60 & & \textbf{81.55} & \textbf{110.15} \\
  & PM$^+$(F+P) & 203 & 32.90 & \textbf{57.20} & & 40.55 & \textbf{62.40} \\
  & LR(F+P) & 203 & 87.20 & 115.20 &  & \textbf{90.70} & 118.60 \\
  & MLP(F+P) & 203 & 108.45 & 142.55 &  & 98.25 & 138.20 \\
  \midrule
  \multirow{10}{*}{\shortstack[l]{Full}} & EM(F) & 348 & 77.00 & \textbf{139.00} &  \multirow{10}{*}{\shortstack[l]{Full}} & \textbf{77.00} & \textbf{139.00} \\
  & PM$^*$(F) & 348 & 82.00 & \textbf{151.00} & & \textbf{82.00} & \textbf{151.00} \\
  & PM$^+$(F) & 348 & 86.00 & \textbf{155.00} & & \textbf{86.00} & \textbf{155.00} \\
  & LR(F) & 348 & \textbf{90.00} & \textbf{152.00} & & \textbf{90.00} & \textbf{152.00} \\
  & MLP(F) & 348 & \textbf{85.00} & \textbf{150.00} & & \textbf{85.00} & \textbf{150.00} \\
  & EM(F+P) & 203 & \textbf{84.00} & \textbf{108.00} & & \textbf{84.00} & \textbf{108.00} \\
  & PM$^*$(F+P) & 203 & \textbf{49.00} & \textbf{74.00} & &  49.00 & 74.00 \\
  & PM$^+$(F+P) & 203 & \textbf{33.00} & 57.00 & & 33.00 & 57.00 \\
  & LR(F+P) & 203 & \textbf{88.00} & \textbf{120.00} &  & 88.00 & \textbf{120.00} \\
  & MLP(F+P) & 203 & \textbf{113.00} & \textbf{145.00} &  & \textbf{113.00} & \textbf{145.00} \\
  \bottomrule
  \end{tabular}
  }
\end{table} 

\subsection{Sampling Impact (RQ3)}
\label{sec:result_RQ3}


To investigate how the mutation sampling rates affect the performance of \name,
we attempt to localise the studied faults using mutants sampled with different
rates. Table~\ref{tab:RQ3} (left side) shows the uniform sampling results with  
rates of 0.1, 0.3, 0.5, and 0.7: all metric values are 
averaged across 20 different samples.
Table~\ref{tab:RQ3} also 
includes the results obtained without sampling (Full). The best results are 
typeset in bold.

As expected, the Full configuration often shows the best performance, followed
by sampling rates of 0.7 and 0.5. Since we expect different mutants to 
contribute different amounts of information to localisation, we do not find it
surprising that sampling rates down to 0.5 show comparable results with
the Full configuration. However, the performance does not degrade at the same
rate as the sampling rate, as can be seen from the results obtained using 
the sampling rate of 0.1.

Since larger methods are likely to produce more mutants, uniform sampling will
effectively sample more mutants for larger methods. We investigate whether 
this is disadvantageous for relatively smaller methods by evaluating stratified
sampling: given the threshold parameter $N$, stratified sampling randomly 
chooses only $N$ mutants from methods with more than $N$ mutants, and chooses
all mutants if their number is below $N$. Table~\ref{tab:RQ3} (right side)
contains the results obtained using stratified mutant sampling with $N \in \{5, 10, 15, 20\}$.
The value in the parenthesis, i.e., "Ratio", is the average ratio of the 
number of mutants sampled by stratified sampling to the number of all mutants.

Compared to the Full configuration, the performance degradation as $N$ 
decreases is notably worse than what has been observed from the results of 
uniform random sampling. However, even with $N = 5$, the sample ratio is 
0.27 on average, higher than the smallest sampling rate for the uniform 
sampling. The comparison suggests that, contrary to our concern for a 
potential bias against smaller methods, stratified sampling is actually 
harmful to \name. One interpretation of the result is that, if we assume that
the location of a fault is a random variable, larger methods are
by definition more likely to contain it.

We answer RQ3 that the impact of mutation sampling is observable but not 
too disruptive. Using uniform sampling, on average 80\%
of the faults ranked at the top without sampling can still be localised
at the top. However, stratified sampling actually harms \name: larger 
methods need to be represented by more mutants.

\section{Discussion}
\label{sec:discussion}

\subsection{Relation with Other FL Techniques}
\label{sec:relation}

\begin{figure}[ht]
  \centering
  \begin{minipage}[t]{.22\textwidth}
    \centering
        \includegraphics[width=\textwidth]{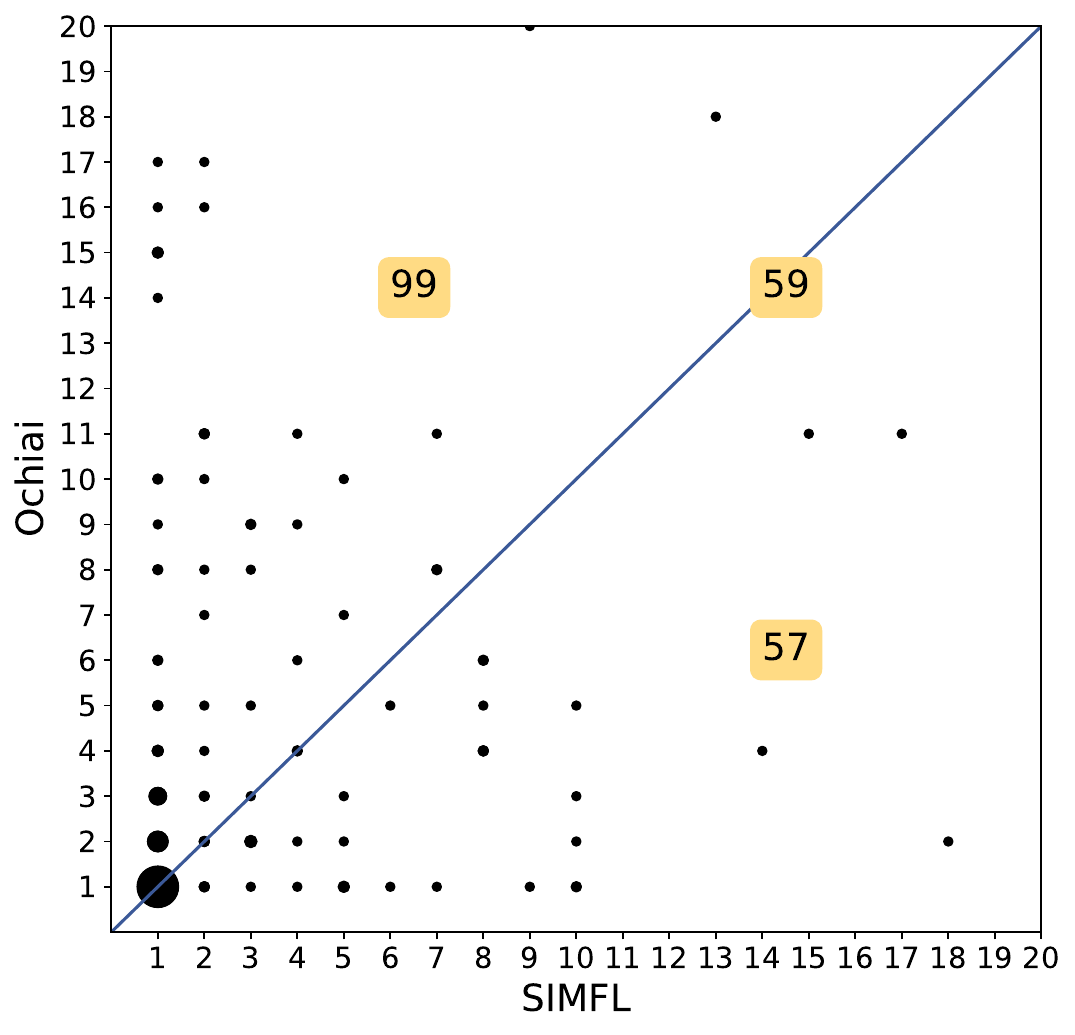}
        \vspace{-1.5em}
        \subcaption{Ochiai ($r = 0.215$)\label{fig:vs_ochiai}}
  \end{minipage}%
  \begin{minipage}[t]{.22\textwidth}
    \centering
        \includegraphics[width=\textwidth]{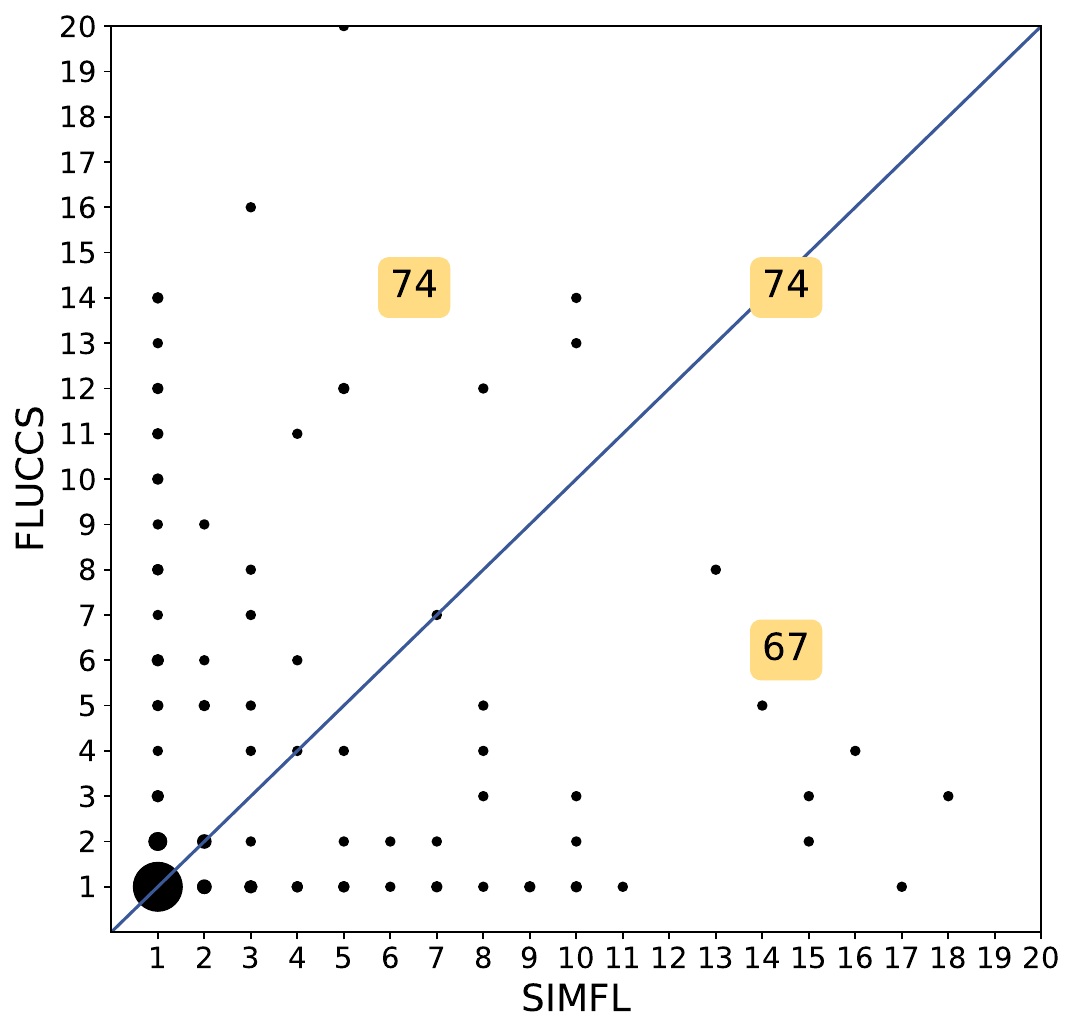}
        \vspace{-1.5em}
        \subcaption{FLUCCS ($r = 0.105$)\label{fig:vs_fluccs}}
  \end{minipage}
  \begin{minipage}[t]{.22\textwidth}
      \centering
      \includegraphics[width=\textwidth]{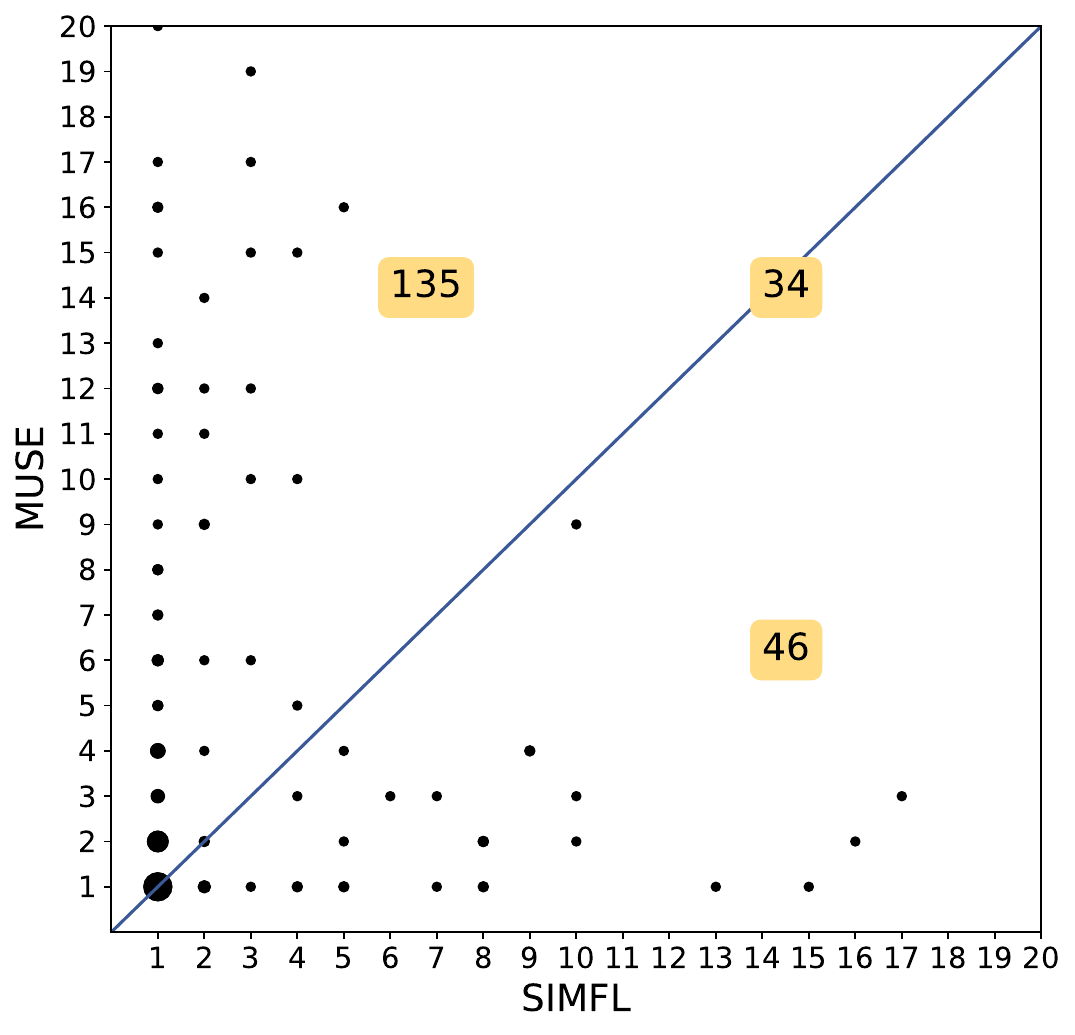}
      \vspace{-1.5em}
      \subcaption{MUSE ($r = 0.072$)\label{fig:vs_muse}}
  \end{minipage}%
  \begin{minipage}[t]{.22\textwidth}
    \centering
      \includegraphics[width=\textwidth]{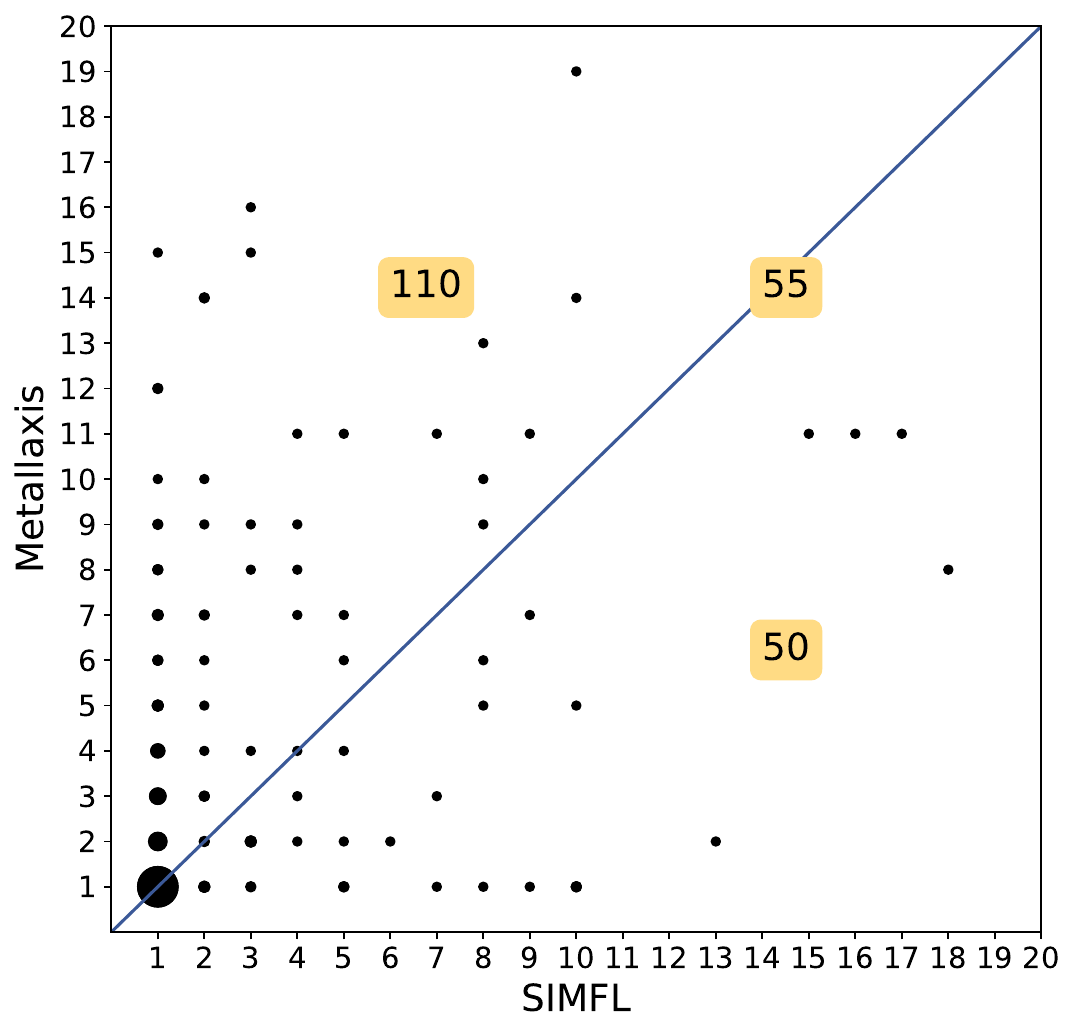}
      \vspace{-1.5em}
      \subcaption{Metallaxis ($r = 0.130$)\label{fig:vs_metallaxis}}
  \end{minipage}
  \caption{Comparison of MLP(F+P) and other FL techniques.
  \label{fig:one_to_one_comparison}}
\end{figure}

Two FL techniques can be complementary to each other if there is little overlap
between faults ranked highly by each technique. 
To investigate whether the contribution of \name is uniquely different from 
others, we investigate how individual faults are ranked differently by \name 
and other FL techniques. \name is represented by the MLP(F+P) model. 
We omit TraPT from this comparison as the individual 
rank information was not available from the paper; DStar is also excluded
as its results are very similar to that of Ochiai.

Figure~\ref{fig:one_to_one_comparison} plots each individual fault according 
to its rank by MLP(F+P) of \name ($x$-axis) and its rank by the other FL 
technique ($y$-axis). Data points on the line $y = x$ represent faults that 
are ranked at the same place by both techniques, whereas the farther from the 
line a point is, the more differently it is ranked by two techniques. Plots
only contain faults that are ranked within the top 20 places by at least one
technique: the size of the dots corresponds to the number of faults plotted
at the location of the dot. The numbers on the $y=x$ line as well as above and 
below the line show the total number of faults that belong to the corresponding 
parts, regardless of being ranked within the top 20 or not. For example, \name 
ranks 135 faults higher than MUSE.
The agreement between two techniques are measured using Pearson correlation 
coefficient ($r$): value 0 implies no correlation and, therefore, no 
agreement, whereas value 1 implies perfect correlation and, therefore, two 
identical rankings.

While there exist dense clusters of points near the top ranks around the 
$y=x$ line, there is no clear relationship between FL techniques. \name
shows low Pearson correlation coefficients against all compared techniques.
Notably, \name is significantly different from two existing MBFL techniques,
MUSE and Metallaxis, suggesting that the way \name captures the relationship
between faults and tests differs significantly from existing MBFL techniques.
\name also ranks the most faults identically to FLUCCS, a technique that uses
multiple SBFL scores as well as code and change metric, suggesting that 
mutation analysis can be a rich source of information for fault localisation.
Overall, the results provide evidence that there exist 
faults that \name can localise much more effectively than the other, and vice 
versa. The complementary nature also suggests the possibility of an effective 
hybridisation of \name and other techniques, as recent work that combine
multiple FL techniques suggest~\cite{zou2019empirical,Xuan:2014kq,Sohn:2017xq}. 
We leave the hybridisation as future work.


\subsection{Test Case Granularity}
\label{sec:granularity}

 
\begin{table}[ht]
  \centering  
  \caption{The result of one-tailed $t$-test between $W3$ and $O3$.
  \label{tab:test_gran}}        
  \scalebox{0.73}{    
    \begin{tabular}{l|l|r|r|r||l|l|r|r|r}
      \toprule
      Model & Project & $W3$ & $O3$ & $p$ & Model & Project & $W3$ & $O3$ & $p$ \\
            &         & mean & mean &     &       &         & mean & mean & \\
      \midrule
      \multirow{5}{*}{\shortstack[l]{EM(F)}} & Lang & 5.5 & 10.0 & \textbf{0.029} & \multirow{5}{*}{\shortstack[l]{EM(F+P)}} & Lang & 6.2 & 7.9 & 0.238 \\
      & Chart & 10.0 & 144.9 & 0.056 & & Chart & 11.4 & 136.1 & 0.086 \\
      & Time & 69.2 & 129.6 & \textbf{0.002} & & Time & 103.7 & 113.7 & 0.320 \\
      & Closure & 167.2 & 322.7 & \textbf{0.000} & & - & - & - & - \\
      & Math & 15.5 & 34.8 & \textbf{0.000} & & Math & 20.5 & 28.5 & 0.056 \\
      \midrule
      \multirow{5}{*}{\shortstack[l]{PM$^*$(F)}} & Lang & 5.1 & 11.9 & \textbf{0.002} & \multirow{5}{*}{\shortstack[l]{PM$^*$(F+P)}} & Lang & 4.7 & 9.8 & \textbf{0.009} \\
      & Chart & 10.0 & 144.9 & 0.056 & & Chart & 96.3 & 88.4 & 0.534 \\
      & Time & 78.4 & 127.6 & \textbf{0.009} & & Time & 20.0 & 120.3 & \textbf{0.000} \\
      & Closure & 175.7 & 327.6 & \textbf{0.000} & & - & - & - & - \\
      & Math & 15.4 & 36.3 & \textbf{0.000} & & Math & 15.2 & 28.2 & \textbf{0.009} \\
      \midrule
      \multirow{5}{*}{\shortstack[l]{PM$^+$(F)}} & Lang & 5.1 & 12.4 & \textbf{0.002} & \multirow{5}{*}{\shortstack[l]{PM$^+$(F+P)}} & Lang & 4.4 & 9.2 & \textbf{0.012} \\
      & Chart & 9.8 & 136.6 & \textbf{0.072} & & Chart & 96.9 & 88.1 & 0.538 \\
      & Time & 78.4 & 127.6 & \textbf{0.009} & & Time & 0.5 & 117.7 & \textbf{0.001} \\
      & Closure & 203.2 & 323.5 & \textbf{0.000} & & - & - & - & - \\
      & Math & 15.5 & 35.8 & \textbf{0.000} & & Math & 11.5 & 27.1 & \textbf{0.010} \\
      \midrule
      \multirow{5}{*}{\shortstack[l]{LR(F)}} & Lang & 4.9 & 13.5 & \textbf{0.000} & \multirow{5}{*}{\shortstack[l]{LR(F+P)}} & Lang & 5.5 & 12.2 & \textbf{0.006} \\
      & Chart & 11.8 & 135.9 & \textbf{0.087} & & Chart & 70.3 & 117.9 & 0.298 \\
      & Time & 78.4 & 127.6 & \textbf{0.009} & & Time & 90.8 & 129.6 & \textbf{0.031} \\
      & Closure & 193.3 & 322.3 & \textbf{0.000} & & - & - & - & - \\
      & Math & 16.5 & 35.3 & \textbf{0.000} & & Math & 17.3 & 31.0 & \textbf{0.003} \\
      \midrule
      \multirow{5}{*}{\shortstack[l]{MLP(F)}} & Lang & 5.4 & 12.9 & \textbf{0.003} & \multirow{5}{*}{\shortstack[l]{MLP(F+P)}} & Lang & 5.9 & 13.3 & \textbf{0.013} \\
      & Chart & 10.8 & 144.9 & \textbf{0.066} & & Chart & 14.6 & 174.2 & \textbf{0.032} \\
      & Time & 62.3 & 137.7 & \textbf{0.000} & & Time & 95.1 & 130.5 & \textbf{0.049} \\
      & Closure & 168.6 & 325.3 & \textbf{0.000} & & - & - & - & - \\
      & Math & 15.6 & 35.8 & \textbf{0.000} & & Math & 17.0 & 40.9 & \textbf{0.000} \\
      \bottomrule
    \end{tabular}    
  }
\end{table}

A common pattern observed in all configurations of \name is that it performs 
the best for Commons Lang. Following Laghari and Demeyer~\cite{Laghari2018aa}, 
we hypothesise that this may be related to the test case granularity: if each 
test case kills mutants that exist in only a few methods, \name can benefit 
from this because failures of each test case will be tightly coupled with a 
few candidate locations.

To investigate the impact of test case granularity, we check whether the number
of the methods that are relevant to failures caused by highly ranked faults is
lower than the number of methods relevant to faults that are not ranked near 
the top. We define a method $m$ to be relevant to the failure of a test case $t$
if $t$ kills a mutant in $m$. A finer granularity test case $t$ is expected to
be relevant to fewer methods. We categorise faults into those ranked in the top
three places (set $W3$), and those that are not (set $O3$), and compare the number
of relevant methods between $W3$ and $O3$.

Table~\ref{tab:test_gran} reports the result of one-tailed $t$-test on the number of 
relevant methods between $W3$ and $O3$: for 33 out of 45 cases, we accept 
the alternative hypothesis that
the mean of $O3$ is significantly greater than $W3$.
In other words, the faults in $W3$ are likely to be revealed by 
test cases with finer-granularity than the faults in $O3$. The test cases of Commons Lang have 
finer-granularity
when compared to other subjects, leading us to conjecture that test
case granularity is why \name performs more effectively against Lang than
others. However, the results also show that \name is not simply reflecting a
one-to-one mapping between methods (mutants) and their unit tests: failing test cases of
Closure kill mutants in 203 methods on average, but PM$^+$(F) can still localise
41 out of 132 faults within the top three places (see Table~\ref{tab:RQ1_total_ranks}).

\subsection{Kill Reason Filtering}
\label{sec:kill_reason_filtering}
\begin{table}[ht]
  \centering  
  \caption{
  The $acc@n$ metric values after filtering mutants based on their kill reasons.\label{tab:kill_reason_filtering}}  
  \scalebox{0.8}{
  \begin{tabular}{l|l|rrr|rrr|rrr}
    \toprule
    \multirow{2}{*}{Model} & Total   & \multicolumn{3}{c|}{Assertion} & \multicolumn{3}{c|}{Timeout} & \multicolumn{3}{c}{Exception} \\   
                          & Studied    & $@1$                           &  $@3$         & $@5$         & $@1$ & $@3$ & $@5$ & $@1$        & $@3$        & $@5$        \\ \midrule
    EM(F)                  & 348        & \textbf{100}                   &  \textbf{163} & \textbf{179} & 21   & 30   & 38   & 53          & 90          & 107         \\
    PM$^*$(F)              & 348        & \textbf{108}                   &  \textbf{185} & \textbf{206} & 23   & 35   & 44   & 60          & 97          & 117         \\
    PM$^+$(F)              & 348        & \textbf{114}                   &  \textbf{183} & \textbf{206} & 22   & 36   & 45   & 61          & 99          & 119         \\
    LR(F)                  & 348        & \textbf{118}                   &  \textbf{181} & \textbf{213} & 40   & 79   & 88   & 66          & 109         & 131         \\
    MLP(F)                 & 348        & \textbf{121}                   &  \textbf{189} & \textbf{210}          & 43   & 76   & 91   & 60          & 106         & 129         \\ \midrule
    EM(F+P)                & 203                          & 72                             &  89           & 96           & 7    & 11   & 16   & 55          & 64          & 68          \\
    PM$^*$(F+P)            & 203        & \textbf{50}                    &  \textbf{76}  & \textbf{90}  & 23   & 42   & 57   & 49          & 71          & 80          \\
    PM$^+$(F+P)            & 203        & \textbf{34}                    &  57           & \textbf{68}  & 20   & 34   & 50   & \textbf{34} & \textbf{59} & \textbf{68} \\
    LR(F+P)                & 203        & \textbf{89}                    &  117          & 128          & 23   & 41   & 51   & 77          & 104         & 115         \\
    MLP(F+P)               & 203        & 112                            &  \textbf{147} & \textbf{160}          & 29   & 50   & 55   & 91          & 120         & 135         \\
    \bottomrule
    \end{tabular}
  }
\vspace{-1em}
\end{table}
A mutated program can cause a test failure due to many different reasons, such 
as assertion (i.e., test oracle) violation, uncaught exception, or timeout. 
All these reasons are normally marked as a kill. While all three reasons do
reveal some dependency between the mutated location and the test outcome
(otherwise the mutant would not be killed), we suspect that different kill
reasons may have varying degrees of importance for fault localisation. 
Assertion violations would imply that the test oracles actually capture
the correct program behaviour. Uncaught exceptions and timeouts, however,
may only show coincidental impacts of the mutation.

Considering the relative importance of different kill reasons, we 
investigate whether filtering out the kill matrix based on the exact reason of 
test failure has any impact on the localisation effectiveness. This is
partly motivated by the use of failure messages by 
TraPT~\cite{li2017transforming}.
We train \name models using one of three kill reasons,
and compare their results to those of models trained
using all three reasons. Kill reasons supported by Major are: assertion 
violations ("Assertion"), timeouts ("Timeout"), and uncaught exceptions ("Exception").

Table~\ref{tab:kill_reason_filtering} shows the results of $acc@n$ metrics for
\name models of three different kill reasons. For all F models, using only mutants 
killed due to the assertion failures shows the best performance in terms 
of $acc@1$ and $acc@3$, adding support to our assumption that
assertion violations reflect test oracles of correct program behaviour better than others.
Timeouts appear to be the weakest signal.


However, for F+P models, the unfiltered original results ("All") often show 
the best performance. This trend reveals a seemingly 
counter-intuitive, yet fundamental intuition about \name: test cases in $\mathbf{T}_f$ 
and $\mathbf{T}_p$ contribute to localisation in different ways. If a test case $t$ is in $\mathbf{T}_f$, all mutants 
killed by $t$ earlier suggest that their locations may contain the fault. 
However, if $t \in \mathbf{T}_p$, all mutants killed by $t$ earlier suggests that their 
locations may \emph{not} contain the fault that is detected by $t' \in \mathbf{T}_f$.
Consequently, kill reason filtering can make the contributions from tests in 
$\mathbf{T}_f$ more precise (i.e., to only reflect real fault detection), but may also 
reduce the total amount of contributions from tests in $\mathbf{T}_p$ because it removes 
potential locations that could have been \emph{excluded} by being associated 
with a test in $\mathbf{T}_p$. This explains why, for F+P models, using only Assertion as
the kill reason cannot dominate the results. Note that the distribution of 
kills between Assertion, Timeout, and Exception is likely not uniform, which we
also think contributes to the mixed results of F+P models, combined with 
program semantics.




\section{Threats to Validity}
\label{sec:threats}

Given the controlled setting for our experiments and the clearly defined
objective measures, there are few threats to the internal validity of our study.
There are some threats to internal validity that are inherent to any mutation
analysis and hard to completely avoid, such as non-determinism caused by mutation 
and equivalent mutants, which have been discussed in Section~\ref{sec:result}.
Similarly, we see few threats to the construct and conclusion validity. The
metrics we used are standard in the fault localisation literature. 
We note that establishing one best technique is not our main goal here and
we would likely need more study subjects for such a comparison to be meaningful.

Rather, the main threat of our study is to its external validity. Even though we
studied five different subjects from the real-world \dfj benchmark
to mitigate this threat, this does not allow us to generalise to many, other
programs and test suite contexts. Still, there was enough variation among the
five subjects for us to identify \name's dependence on the granularity of the
test cases.

\section{Related Work}
\label{sec:related_work}

A number of Mutation Based Fault Localisation techniques have been proposed
in the literature. Metallaxis uses SBFL-like formulas to measure the similarity 
between failure patterns of the actual fault and 
mutants~\cite{Papadakis:2015sf,Papadakis:2012fk}. MUSE~\cite{Moon:2014ly}, 
and its variation MUSEUM~\cite{Hong:2017qy}, depend on two principles: first, 
if we mutate already faulty parts of the program, it is unlikely that we will 
observe more failing test cases, and we may even observe partial fixes, and 
second, if we mutate non-faulty parts, tests that used to fail are now likely
to fail. MUSE and MUSEUM define their suspiciousness scores using the
ratios of fail-become-pass and pass-become-fail tests. 
TraPT is similar to MUSE and MUSEUM in nature, but transforms both the
output messages of failing tests, to distinguish different types of exceptions,
and the test code itself, to prevent early program termination due to the 
assertion violation that precludes collecting information of other assertions
~\cite{li2017transforming}. All existing MBFL techniques mutate the faulty
program once testing is finished. In contrast, \name allows the mutation 
analysis to be performed ahead of time.


\name was initially formulated based on Bayesian analysis to infer likely fault 
locations given test information. In the context of fault localisation, 
Abreau et al.~\cite{Abreu:2009qy} have introduced \textsc{Barinel}, an SBFL technique that 
adopts Bayesian reasoning to generate candidate sets of multiple fault
locations. To the best of our knowledge, \name is the first MBFL technique
that uses Bayesian inference as well as other statistical inference techniques.
While \name also uses dynamic information from mutation, 
the mutation analysis can be performed ahead-of-time, which allows the cost 
to be amortised over multiple development iterations, and provides faster 
feedback.

\section{Conclusion}
\label{sec:conclusion}

This paper introduces \name, a Mutation Based Fault Localisation (MBFL)
technique that allows users to perform the mutation analysis in advance,
before the actual failure is observed. \name relies on statistical inference
techniques to train predictive models that can be used with the actual
failure information. This allows us to use the concrete 
and precise dependencies between source code and test cases for fault 
localisation, without having to expend the large cost of mutation analysis 
when failures are observed. We have empirically evaluated \name 
using real-world faults from \dfj benchmark. \name can localise 
113 faults at the top, and is capable of retaining 80\% of its localisation
accuracy at the top when we sample only 10\% of all generated mutants.


\section{Acknowledgement}
\label{sec:ack}
Authors would like to thank Mike Papadakis for the insightful discussion about
mutation based fault localisation. Jinhan Kim, Gabin An, and Shin Yoo have been
supported by National Research Foundation of Korea (NRF) Grant
(NRF-2020R1A2C1013629), Institute for Information \& communications Technology
Promotion grant funded by the Korean government (MSIT) (No.2021-0-01001), and
Samsung Electronics (Grant No. IO201210-07969-01). Robert Feldt acknowledges
support from two Science Council projects (Vetenskapsrådet, contract ids
2015-04913-9 and 2020-05272).

\bibliographystyle{IEEEtran}
\balance
\bibliography{newref}
\end{document}